\documentclass[aps,showpacs]{revtex4}
\usepackage{graphicx}
\usepackage{dcolumn}
\usepackage{bm}
\usepackage{graphicx}
\usepackage{subfigure}

\begin{document}

\title{Superluminal tunneling of microwaves in smoothly varying transmission lines}
\author{A. B. Shvartsburg}
\affiliation{Joint Institute of High Temperatures Russian Academy of Sciences, Moscow, Russian Federation}

\author{M. Marklund}
\affiliation{Department of Physics, Ume{\aa} University, SE--901 87 Ume{\aa},
Sweden}

\author{G. Brodin}
\affiliation{Department of Physics, Ume{\aa} University, SE--901 87 Ume{\aa},
Sweden}

\author{L. Stenflo}
\affiliation{Department of Physics, Ume{\aa} University, SE--901 87 Ume{\aa},
Sweden}

\begin{abstract}
Tunneling of microwaves through a smooth barrier in a
transmission line is considered. In contrast to standard wave barriers, we
study the case where the dielectric permittivity is positive, and the
barrier is caused by the inhomogeneous dielectric profile. It is found that
reflectionless, superluminal tunneling can take place for waves with a
finite spectral width. The consequences of these findings are discussed, and
an experimental setup testing our predictions is proposed.
\end{abstract}
\maketitle

\section{Introduction}

Tunneling is a fundamental process related to the dynamics of various kinds
of waves. This phenomenon was already pointed out in Gamow's famous work 
\cite{Gamow1928} on nuclear $\alpha $-decay, where the probability of
penetration of $\alpha $-particles with energy $E$ through a potential
barrier with height $U_{0}$ (with $U_{0}>E$ ) was found to be exponentially
small, but finite. However, the Gurney and Condon attempt \cite{Condon1928}
to find the velocity $v_{t}$ and the transit time $\tau $ of such a
tunneling revealed a basic problem of the theory; how should one define
these quantities in the ''classically forbidden zone'' ($U_{0}>E$), where
both the values of $v_{t}$ and $\tau $ are imaginary?

Three decades later the interest in this problem was renewed in Hartmann's
paper \cite{Hartmann1962}, where the time of tunneling of a particle with
energy $E$ through a barrier was determined via the phase of the complex
barrier transmission function $T=\left| T\right| \exp (i\varphi _{t})$,
which gives the tunneling time 
\begin{equation}
\tau =\hbar \frac{\partial \varphi _{t}}{\partial E}  \label{EQ-time1}
\end{equation}
where $2\pi \hbar $ is Planck's constant. Using the wellknown expressions
for the transmission function of a rectangular barrier with width $d$,
Hartmann \cite{Hartmann1962} showed that for the case of a thick barrier ($%
\left| \mathbf{p}\right| d\gg \hbar $, where $\mathbf{p}$ is the particle
momentum), the time $\tau $ becomes independent of $d$, and thus for a
sufficiently thick barrier the tunneling speed $v_{t}$ can reach the
superluminal velocity $v_{t}>c$. This conclusion, referred to in the
literature as the Hartmann paradox, can be deduced from standard formula
given in many textbooks. It has stimulated a hot debate which is still going
on \cite{Ciao1997,Mugnai2000,Buttiker2003,Li2001}. However, a direct
measurement of the electron tunneling time through a quantum barrier has
proved to be an intricate task. The idea to verify Hartmann's conclusion by
means of the classical effect of electromagnetic (EM) wave tunneling through
macroscopic wave barriers was then proposed. Its basis was the formal
similarity between the stationary Schr\"{o}dinger equation and the usual
wave equation, describing both propagating and evanescent EM modes in
continuous media.

This idea gave rise to several attempts to examine the tunneling times of EM
waves in microwave and optical ranges. Thus, radio wave experiments with
tunneling of the $TE_{01}$ mode in an ''undersized'' metallic waveguide \cite
{Olkhovsky2004}, were by some authors interpreted in favor of the concept of
superluminal phase times for tunneling EM waves. These conclusions were
reached by means of the expression \cite{Buttiker1988} 
\begin{equation}
\tau _{B}=\sqrt{\left( \frac{\partial \varphi _{t}}{\partial \omega }\right)
^{2}+\left( \frac{\partial \ln \left| T\right| }{\partial \omega }\right)
^{2}}  \label{EQ-timeB}
\end{equation}
generalizing Eq. (\ref{EQ-time1}) to wave tunneling through a barrier
characterized by a complex transmission coefficient. However, measurements
of the group delay time for the fundamental $TE_{01}$ mode near the cut-off
frequency had shown that each of the quantities $\tau $ (\ref{EQ-time1}) and 
$\tau _{B}$ (\ref{EQ-timeB}) correspond to experiments only at some
restricted frequency ranges. Tunneling through such homogeneous opaque
barriers is accompanied by strong reflections, attenuation and reshaping of
the transmitted signal. The destructive interference between the incident
and reflected parts of the pulse was considered in Ref. \cite{Steinberg1993}
as a mechanism for suppressing the tail of the transmitted pulse, causing a
superluminal motion of the pulse peak.

Another approach to this problem is based on the use of a heterogeneous
photonic barrier, characterized by a curvilinear profile of the dielectric
susceptibility $\varepsilon (z)$ across the barrier \cite{Shvartsburg2005}.
Propagation of waves in this geometry is characterized by the following
inhomogeneity induced phenomena:

\begin{description}
\item  a) The appearance of an easily controlled cut-off frequency $\Omega $%
, depending on the gradient and curvature of the profile $\varepsilon (z)$.

\item  b) Formation of a tunneling regime for waves with frequency $\omega
<\Omega $ in dielectric media with $\varepsilon >0$ and $d\varepsilon /dz<0$.

\item  c) Reflectionless (non-attenuative) tunneling providing 100\%
transfer of wave energy through a barrier for some frequency $\omega
_{c}<\Omega $.
\end{description}

These properties were examined in Refs. \cite
{Shvartsburg2005,Shvartsburg2007} for thin dielectric nanofilms,
characterized by a smooth spatial variation of $\varepsilon (z)$ at the
subwavelength nanoscale. Access to such films, being available due to recent
progress in nanotechnology \cite{Nanocoatings}, remains, however, a
challenging technological problem.

The purpose of the present paper is to unite the advantages of both
microwave and optical tunneling systems, to enable measurements of
substantial phase shifts of $\mathrm{GHz}$ microwaves in the regime of
non-attenuative tunneling through heterogeneous dielectric layerss,
characterized, unlike the optical system, by cm-scales for the
inhomogeneity. A candidate for such a device is the coaxial transmission
line (TL), using the fundamental $TEM$ mode. This TL is chosen due to the
following salient features, resembling the properties of EM waves in free
space \cite{Wangsness1986}:

\begin{description}
\item  a) The velocity of the $TEM$ mode is known to coincide with the free
space light velocity $c$.

\item  b) The polarization structure of the $TEM$ mode contains only the
transverse field components of $\mathbf{E}$ and $\mathbf{H}$, and not any
longitudinal component.

\item  c) The $TEM$ mode does not possess any frequency dispersion.
\end{description}

The organization of the present paper is as follows: Tunneling of the $TEM$
mode through a segment of a coaxial TL, containing a diaphragm made from a
gradient dielectric metamaterial is considered in section 2. Conditions for
reflectionless tunneling of a TEM wave through this diaphragm, accompanied
by a substantial phase shift, are examined in section 3. The parameters of
an interferometric scheme for measurements of this phase shift are presented
in section 4. Finally, a brief discussion of the results is given in the
Conclusion (Section 5).

\section{An exactly solvable model for an inhomogeneous transmission line}

Let us consider a coaxial waveguide permitting EM wave propagation between
two infinitely conducting cylinders with radii $b$ and $a$ ($b>a$) shown in
Fig 1. The fundamental $TEM$ mode is assumed to propagate along this
waveguide ($z$-direction). The regions $z\leq 0$ (region 0) and $z\geq d$
(region 2) are assumed to be vacuum, whereas region 1 ($0\leq z\leq d$) is
filled with an inhomogeneous dielectric layer, fabricated from a
metamaterial, where the dielectric susceptibility is varying such that 
\begin{equation}
\varepsilon _{1}=\varepsilon _{m}W^{2}(z),\;  \label{Eq:epsilon}
\end{equation}
where $W^{2}(z)$ (with $W(0)=1$ and$\;W(d)=1$) is a dimensionless positive
function and $\varepsilon _{m}$ is the maximum value of $\varepsilon $
reached at the ends $z=0$ and $z=d$ of the inhomogeneous region. The spatial
variations of the current $I$ and the voltage $V$ along this transmission
line are governed by the wellknown equations \cite{Wangsness1986} 
\begin{equation}
\frac{\partial I}{\partial z}+C\frac{\partial V}{\partial t}=0\text{ \ and }%
\frac{\partial V}{\partial z}+M\frac{\partial I}{\partial t}=0
\label{Eq:Transmission-line}
\end{equation}
Here $M$ and $C$ are the inductance and capacitance per unit length. Losses
are ignored. The values $C_{n}$ and $M_{n}$ are for each region are 
\begin{equation}
C_{n}=\frac{2\pi \varepsilon _{0}\varepsilon _{n}}{\ln (\frac{b}{a})}\;\ \ 
\text{and }M_{n}=\frac{\mu _{0}\mu _{n}}{2\pi }\ln \left( \frac{b}{a}\right)
,\;  \label{eq:capacitance-inductance}
\end{equation}
where $n=0,1,2$ and where $\varepsilon _{0}$ and $\mu _{0}$ are the
dielectric susceptibility and magnetic permeability of vacuum, respectively.
To solve the system (\ref{Eq:Transmission-line}) in the segment $n=1$, we
introduce the generating function $\psi $ defined by 
\begin{equation}
V=-M_{1}\frac{\partial \psi }{\partial t}\text{ \ \ and \ }\;I=\frac{%
\partial \psi }{\partial z}  \label{Eq:generating-func}
\end{equation}
Next, considering a harmonic dependence $\propto \exp (-i\omega t)$ we
obtain 
\begin{equation}
\frac{\partial ^{2}\psi }{\partial z^{2}}+\frac{W_{n}^{2}(z)}{v_{n}^{2}}%
\omega ^{2}\psi =0
\end{equation}
where $v_{n}^{2}=1/M_{n}C_{n}$ such that $v_{0}=v_{2}=c$ and $W_{0}=$ $%
W_{2}=1$. Let us now consider a model where $W_{1}^{2}(z)$ contains two free
parameters $L_{1}$ and $L_{2}$ which can be considered as the characteristic
spatial scales of the inhomogeneity (see Fig 2) 
\begin{equation}
W_{1}^{2}=\frac{1}{\left( 1+\frac{z}{L_{1}}-\frac{z^{2}}{L_{2}^{2}}\right)
^{2}},  \label{Eq:inhom-profile}
\end{equation}
These parameters can be expressed in terms of the minimum value of the
dielectric susceptibility, $\varepsilon _{\min }$, reached in region 1 due
to the profile (\ref{Eq:inhom-profile}). Introducing the ratio $%
y=L_{2}/2L_{1}$ we find \cite{SSW1997} 
\begin{equation}
y=\sqrt{\frac{\varepsilon _{m}}{\varepsilon _{\min }}-1}\text{, }\;L_{2}=%
\frac{d}{2y}\text{ \ and }\;L_{1}=\frac{d}{4y^{2}}  \label{Eq:Constants}
\end{equation}

Next we introduce the phase path length $\eta (z)=\int_{0}^{z}W(z_{1})dz_{1}$%
, and the new function $\psi =F/\sqrt{W}$. This leads to the resulting
equation 
\begin{equation}
\frac{\partial ^{2}F}{\partial \eta ^{2}}-q^{2}F=0  \label{Eq:final-variable}
\end{equation}
with $q^{2}=(\omega ^{2}/v_{1}^{2})(u^{2}-1)$ and $u=\Omega /\omega $ where $%
u>1$. Here $\Omega $ is a cut-off frequency arising due to the inhomogeneous
profile $W_{1}^{2}(z)$. It is 
\begin{equation}
\Omega =\frac{2yv_{1}(1+y^{2})^{1/2}}{d}\;\text{where }v_{1}=\frac{c}{\sqrt{%
\varepsilon _{m}}}  \label{Eq:cut-off}
\end{equation}
Unlike the evanescent modes in Lorentz media [17], characterized by a local
frequency dispersion of natural media, Eq. (11) shows the possibility of
tunneling of LF waves ($\omega <\Omega $ ) in a metamaterial with nonlocal
dispersion. The solution for $F$ is a sum of a forward and a backward
propagating wave, i.e. $F=e^{-}+Qe^{+}$ such that 
\begin{equation}
\psi =\frac{A\left( e^{-}+Qe^{+}\right) }{\sqrt{W}}  \label{Eq:psi-solution}
\end{equation}
where $e^{-}=\exp (-q\eta )$ and $e^{+}=\exp (q\eta )$ and $A$ is a
normalization constant. Using Eqs. (\ref{Eq:generating-func}) the voltage
and current are 
\begin{equation}
V_{1}=\frac{i\omega M}{v_{1}}A\left[ \frac{e^{-}+Qe^{+}}{\sqrt{W}}\right] 
\text{ and}\;I_{1}=Aq\sqrt{W}\left[ \frac{1}{qL_{2}}\left( y-\frac{z}{L_{1}}%
\right) \left( e^{-}+Qe^{+}\right) -\left( e^{-}-Qe^{+}\right) \right] 
\label{Eq: I-V-solutions}
\end{equation}
respectively. The continuity of the current and voltage at $z=0$ gives us
the reflection coefficient 
\begin{equation}
R_{v}=\frac{1+\chi }{1-\chi }  \label{Eq:reflection}
\end{equation}
with 
\begin{equation}
\chi =\frac{iZ_{0}N}{Z_{1}}\left[ \frac{1}{2qL_{1}}-\frac{1-Q}{1+Q}\right] 
\label{eq:x-def}
\end{equation}
where $N=\sqrt{u^{2}-1}$ whereas $Z_{0}$ and $Z_{1}$ are the impedances of
regions 0, 2 and 1, respectively, i.e. 
\begin{equation}
Z_{0}=Z_{2}=\sqrt{\frac{M_{0}}{C_{0}}}\text{ and}\;Z_{1}=\frac{Z_{0}}{\sqrt{%
\varepsilon _{m}}}
\end{equation}
The boundary condition at $z=d$ gives 
\begin{equation}
Q=\frac{\exp (2q\eta _{0})\left[ \beta N+\frac{\gamma }{2}+i\right] }{\left[
\beta N-\frac{\gamma }{2}-i\right] }  \label{Eq:Boundary-1}
\end{equation}
where 
\begin{eqnarray}
\beta  &=&\frac{Z_{0}}{Z_{1}}=\sqrt{\varepsilon _{m}},\;\gamma =\frac{N\beta 
}{\omega L_{1}}=\frac{2\beta uy}{\sqrt{1+y^{2}}}\;,  \nonumber \\
\eta _{0} &=&\eta _{0}(d)=L_{2}/(1+y^{2})^{1/2},l_{0}=\ln (y_{+}/y_{-})\text{
and }y_{\pm }=(1+y^{2})^{1/2}\pm y  \label{eq:def2}
\end{eqnarray}

Thus, we have found a spatial structure of the $E$ and $H$ components of the
evanescent $TEM$ mode. Unlike the traditional concept of a homogeneous
coaxial TL, which is known to have no cut-off frequency for the fundamental $%
TEM$ mode \cite{Wangsness1986}, our inhomogeneous region for a coaxial
waveguide has been shown to provide a cut-off frequency for this mode
without any deformation of the coaxial cross-section. The drastic
consequences of this inhomogeneity induced effect for reflectance and
transmission of the $TEM$ mode are discussed below.

\section{Windows of transparency for the evanescent wave (reflectionless
tunneling of the TEM mode)}

In order to find the reflectance of the gradient wave barrier, we substitute
the formula (\ref{Eq:Boundary-1}) for $Q$, into Eqs. (\ref{Eq:reflection})-(%
\ref{eq:x-def}). This yields the explicit expression for the complex
reflection coefficient 
\begin{equation}
R_{v}=\frac{\tanh (q\eta _{0})\left( 1+\frac{\gamma ^{2}}{4}+\beta
^{2}N^{2}\right) -\gamma \beta N}{\tanh (q\eta _{0})\left( 1-\frac{\gamma
^{2}}{4}-\beta ^{2}N^{2}\right) +\gamma \beta N+i(2\beta N-\gamma \tanh
(q\eta _{0}))}.  \label{Eq:reflection-a}
\end{equation}
Substitution of (\ref{Eq:reflection-a}) into the continuity condition for
the voltage at $z=0$: $V_{in}(1+R_{v})=i\omega _{c}^{-1}A(1+Q)$, where $%
V_{in}$ is the voltage of the incident wave, determines the normalization
constant $A$. Then, substitution of the constant $A$ into (\ref{Eq:
I-V-solutions}) gives the complex transmission function $T=\left| T\right|
\exp (i\varphi _{t})$, i.e. 
\begin{equation}
\left| T\right| =\sqrt{1-\left| R_{v}\right| ^{2}}  \label{Eq:transmittance}
\end{equation}
with 
\begin{equation}
\varphi _{t}=\tan ^{-1}\left\{ \frac{\tanh (q\eta _{0})\left( 1-\frac{\gamma
^{2}}{4}-\beta ^{2}N^{2}\right) +\gamma \beta N}{2\beta N-\gamma \tanh
(q\eta _{0})}\right\}   \label{Eq:phase}
\end{equation}
The formula (\ref{Eq:reflection-a}) gives the reflectance of a single
inhomogeneous layer. If the wave is tunneling through $m$ ($m\geq 1$)
neighboring and identical layers, the total reflection coefficient $R_{v}$
and phase $\varphi _{t}$ are found from (\ref{Eq:reflection-a}) and (\ref
{Eq:phase}) by the replacement

\begin{equation}
\tanh (q\eta _{0})\rightarrow \tanh (mq\eta _{0})  \label{Eq:substitution}
\end{equation}
We note that the reflection coefficient $R_{v}$ can be zero, which, for a
series of $m$ layers, corresponds to the condition 
\begin{equation}
\tanh (mq\eta _{0})=\frac{\gamma N\sqrt{\varepsilon _{m}}}{\left( 1+\frac{%
\gamma ^{2}}{4}+\varepsilon _{m}N^{2}\right) }  \label{Eq:zero-R-m}
\end{equation}
The parameters $\gamma $ and $\eta _{0}$ are defined by (\ref{eq:def2}). Eq.
(\ref{Eq:zero-R-m}) determines the normalized critical frequency $u_{c}=\Omega
/\omega_{c} $ that results in reflectionless tunneling.

To optimize the parameter values for an inhomogeneous reflectionless
barrier, one has to choose the maximum and minimum values of the dielectric
susceptibility, $\varepsilon _{m}$ and $\varepsilon _{\min }$ respectively,
and the number of barriers $m$. The next steps are:

\begin{enumerate}
\item  Calculation of the values $y$ from (\ref{Eq:Constants}) and $v_{1}$
from (\ref{Eq:cut-off}).

\item  Solution of Eq. (\ref{Eq:zero-R-m}) for the unknown normalized critical
frequency $u_{c}=\Omega /\omega _{c}$ corresponding to the above values for $%
y$ and $v_{1}$,

\item  Determination of the product $\omega _{c}d$ from the condition
following from (\ref{Eq:cut-off})
\end{enumerate}

\begin{equation}
\omega _{c}d=\frac{2yv_{1}(1+y^{2})^{1/2}}{u_{c}}=K  \label{Eq24}
\end{equation}
As a numerical example of such reflectionless tunneling one can consider the
values $y_{2}=1/3$, $\varepsilon _{m}=4.912$, $m=2$; in this case the
solution of Eq. (\ref{Eq24}) is $u_{c}=1.0205$ and the constant $K=1.761010$ 
\textrm{cm/s}. Then for a frequency $\omega _{c}/2\pi =1\mathrm{GHz}$ ($%
\lambda =30\mathrm{cm}$) we obtain $d=2.8\mathrm{cm}\ll \lambda $. We note
that the condition for reflectionless tunneling (complete transmission $%
R_{v}=0$, $\left| T\right| =1$) can be fulfilled for different frequencies $%
\omega $ and lengths $d$, linked by Eq. (\ref{Eq:zero-R-m}). An example of
the transmission spectrum for reflectionless tunneling for the normalized
frequencies $u$ is presented in Fig 3a. The transmission is almost complete, 
$\left| T\right| ^{2}>0.95$ in a finite spectral range ($1.02<u<1.25$),
forming a transparency window for the tunneling mode.

\section{Phase effects in the transmitted mode: superluminal tunneling?}

The phase shift $\varphi _{t}$ of the tunneling mode (\ref{Eq:phase}) is
depicted in Fig 3b, as a function of a normalized frequency $u$, for fixed
parameters $\varepsilon _{m}$ and $\varepsilon _{\min }$. For the
reflectionless tunneling under discussion this shift is $\varphi _{t}=1.6$ $%
\mathrm{rad}$. This graph can be used for different frequencies $\omega $
and lengths $d$, linked by a constant product $\omega d$. We notice that
this constant product $\omega d$ determines the phase shift $\varphi
_{0}=\omega d/c$ of a wave with frequency $\omega $ accumulated along the
path $d$ in free space.

To measure the phase shift $\varphi _{t}$ of the tunneling wave, and
comparing it with $\varphi _{0}=\omega d/c$, one can use the
interferometer-like scheme depicted in Fig. 4. The $TEM$ mode, produced by a
generator G, is splitted between two similar coaxial transmission lines. The
space between the cylinders in TL 1 is empty, whereas the second one
contains the inhomogeneous barrier that has been discussed above. After
passage through these arms, the two waves are interfering, and the relative
phase shift $\Delta \varphi =$ $\varphi _{t}-\varphi _{0}$ can be measured.
Let us write 
\begin{equation}
\varphi _{t}=\varphi _{0}\left( 1+\frac{\Delta \varphi }{\varphi _{0}}%
\right)   \label{eq:phase-diff}
\end{equation}
and introduce the phase-time delay $\mathcal{T}$ and the relevant velocity $%
v_{t}$, linked by the condition $v_{t}\mathcal{T}=D=t_{0}c$. Then, putting $%
\varphi _{0}=\omega t_{0}$ and $\varphi _{t}=\omega t_{0}v_{t}/c$, we find 
\begin{equation}
\frac{v_{t}}{c}=1+\frac{\Delta \varphi }{\varphi _{0}},\;\mathcal{T}=\frac{%
t_{0}}{1+\frac{\Delta \varphi }{\varphi _{0}}}\text{ and }\;t_{0}=\frac{d}{c}
\label{eq:super-lum}
\end{equation}
Thus in the case $\Delta \varphi >0$ ($\Delta \varphi <0$) the scheme
discussed will result either in a subluminal ($v_{t}<c$, $\mathcal{T}>t_{0}$%
) or a superluminal ($v_{t}>c$, $\mathcal{T}<t_{0}$) regime. For the figures
related to the above mentioned reflectionless tunneling ($m=2$, $D=2d$) one
can find from (\ref{Eq:phase}), $\Delta \varphi =0.43\mathrm{rad}$, which
indicates a superluminal propagation of the evanescent mode with $v_{t}=1.35c
$ and $\mathcal{T}=0.74t_{0}$.  Unlike the phase velocity of propagating
wave $v_{\phi }=\omega /k$, characterizing the continuous accumulation of
phase (with the wavenumber $k$ real), we consider the velocity connected
with the phase of evanescent wave, which is not associated with its
wavenumber. Some salient features of such a superluminal tunneling are:

\begin{enumerate}
\item  Unlike the case with the Hartmann geometry \cite{Hartmann1962}, no
thick barriers and phase saturation are needed in the present case. 

\item  The power flow of the $TEM$ mode is completely transmitted by means
of the evanescent wave.

\item  The energy density of the evanescent mode inside the barrier $%
(\varepsilon \left| E\right| ^{2}+\mu \left| H\right| ^{2})/2$ remains
positive in each cross-section of the barrier.
\end{enumerate}

It is interesting to evaluate the group time delay for this type of
tunneling by means of equations (\ref{EQ-time1}) and (\ref{EQ-timeB}). Since
we consider the case of reflectionless tunneling ($\left| T\right| =1$), the
second term in (\ref{EQ-timeB}) vanishes, and we can use (\ref{EQ-time1}).
Presenting the derivative $\partial \varphi _{t}/\partial \omega $ from (\ref
{Eq:phase})  we can write 
\begin{equation}
\tau =\frac{d}{v(u)}\text{ and}\;v(u)=-\frac{K}{u\frac{\partial \varphi _{t}%
}{\partial u}}
\end{equation}
The constant $K$ is given in (\ref{Eq24}). Thus, the time $\tau $ is
proportional to the length $d$ when the normalized frequency $u$ is given.
For the example discussed, i.e. $\omega =1\mathrm{GHz}$ and $d=2.8\mathrm{cm}
$, the group delay is negative, or $\tau =-0.6\mathrm{ns}$. Here $\left|
\tau \right| $ is subluminal, i.e. $\left| \tau \right| >D/c=0.185\mathrm{ns}
$. By considering the frequency $\omega =4\mathrm{GHz}$, ($d=0.70\mathrm{cm}$%
) keeping the product $\omega d\mathrm{\ }$constant one finds $\tau =-015%
\mathrm{ns}$, which implies a superluminal value $\left| \tau \right| <D/c$.

One has to emphasize that these tunneling wave phenomena do not violate the
Einsteinian causality related to travelling waves.

\section{Conclusions}

In conclusion, we stress the following points:

\begin{enumerate}
\item  The possibility of reflectionless tunneling of microwave TEM mode in
a coaxial waveguide with a gradient profile (i.e.\ gradient media) given by $%
\varepsilon (z)$ has here been examined in the framework of an exactly
solvable model. This model describes the effects of heterogeneity-induced
dispersion and the controlled formation of a cut-off frequency $\Omega $ in
the wanted spectral range. Unlike evanescent waves in Lorentz media with a
natural local dispersion \cite{Buttiker2003}, the tunneling of LF modes ($%
\omega <\Omega $) arises in gradient media due to non-local dispersion.

\item  In contrast to the traditional concept of tunneling in media with $%
\varepsilon <0$, another mechanism of tunneling, with $\varepsilon >0$ but $%
\nabla \varepsilon <0$, has been considered here. The possibility of a
superluminal phase shift, arising in subwavelength wave barriers (beyond
Hartmann's condition $|\mathbf{p}d\gg \hbar |$), is demonstrated. An
experimental setup, illustrating such tunneling, is suggested.

\item  The results, obtained here for microwaves, can easily be generalized
to other types of EM waves, keeping the condition of a constant product $%
\omega d$, with the other parameters of tunneling media being the same.
Moreover, the tunneling effects discussed here seem to be rather general,
occurring for different types of waves satisfying heterogeneous wave
equations \cite{Nimtz2003} for media with continuous spatial variations of
its parameters.
\end{enumerate}

\newpage {\large {Figure Captions} }

\bigskip 

Fig 1:  A schamtic picture of a coaxial waveguide with a gradient wave
barrier in segment 1 between the empty segments 0 and 2. A spatial 
distribution of the normalized impedance $W^{2}(z)$
plotted as a function of $z$ \ ($y_{2}=1/3$) can also be seen.

\bigskip 

\bigskip 

Fig 2: The transmittance spectra for the TEM mode tunneling through two \
wave barriers in the coaxial waveguide ($\varepsilon _{m}=4.912$,  $%
\varepsilon _{\min }=3.69$, $m=2$, $y_{2}=1/3$);  

a) The spectral window of transparency in the range  $1<u<10$ and $1.02 < u < 1.25$. 

b) The phase spectrum of the tunneling mode in the range  $1<u<10$ and $1.02 < u < 1.25$. 

\bigskip 

Fig. 3: An interferometer like scheme for measurement of phase shift of the
tunneling mode. G is the generator of the TEM mode, 1 is an arm containing
two  adjoining gradient impedance layers, 2 is an arm containing the empty
coaxial,  P is a phase  detecting element. 

\newpage

\begin{figure}
\includegraphics[width=0.98\columnwidth]{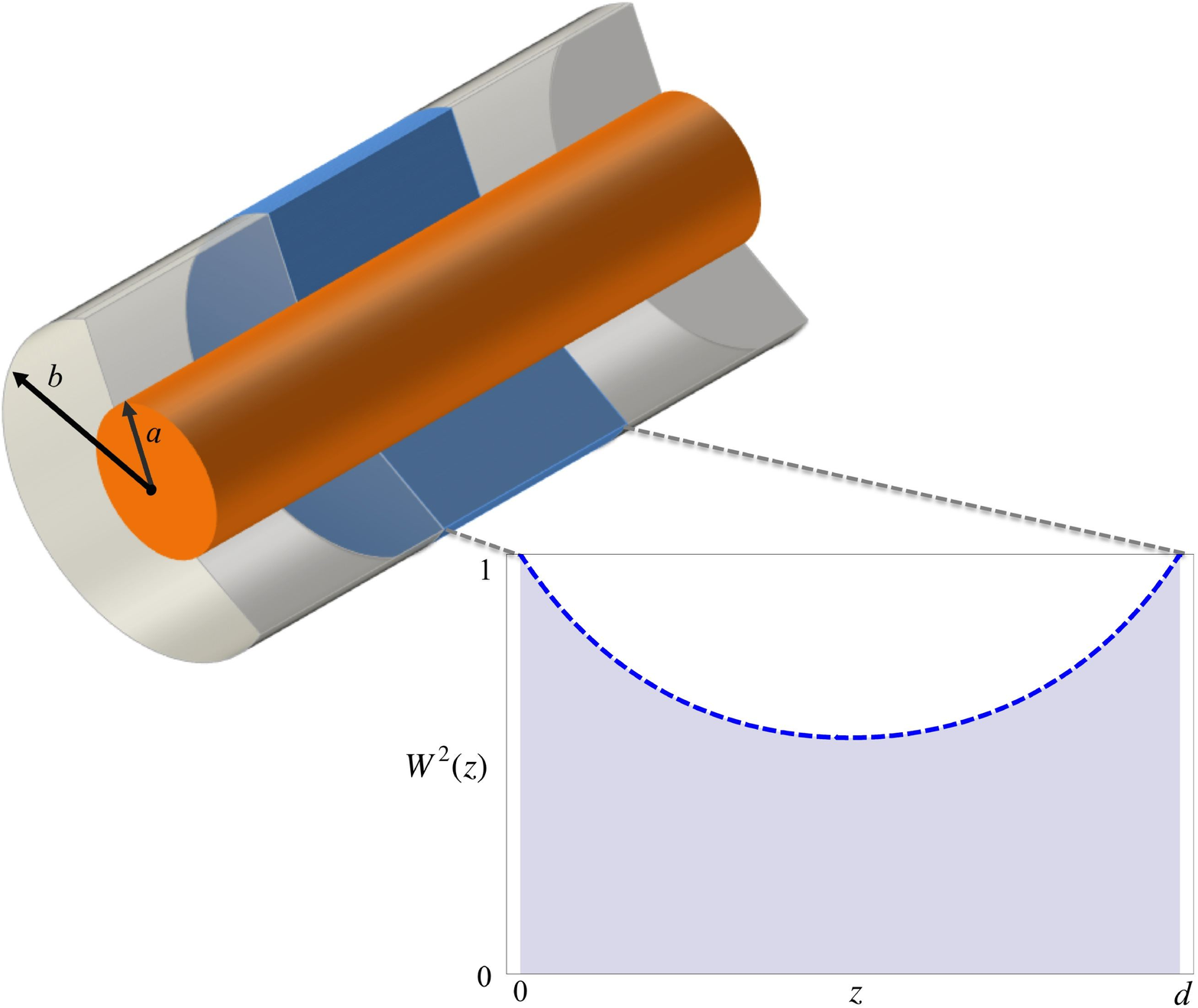}
\caption{}
\end{figure}

\newpage

\begin{figure}
\subfigure[]{\includegraphics[width=0.98\columnwidth]{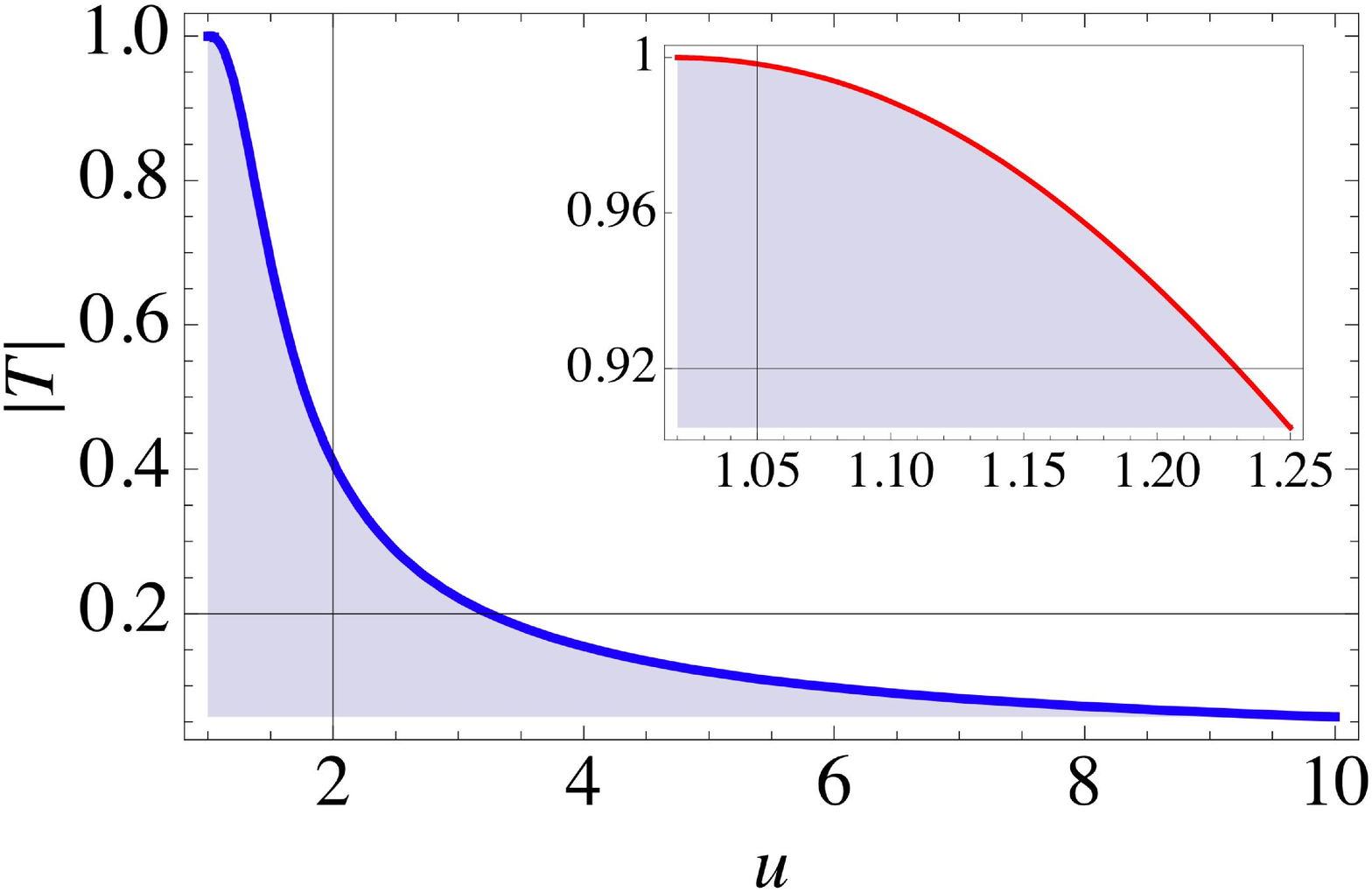}}
\subfigure[]{\includegraphics[width=0.98\columnwidth]{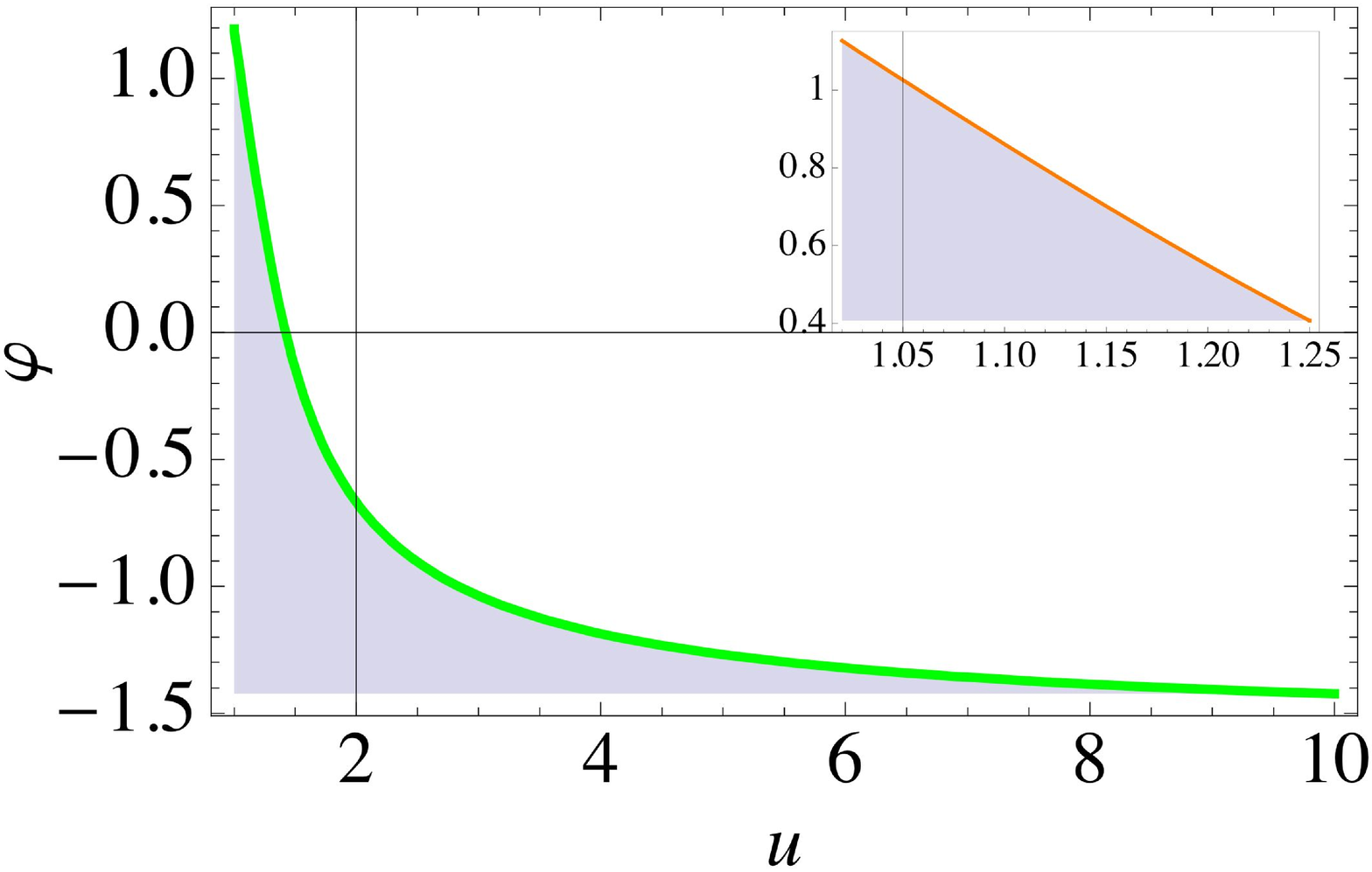}}
\caption{}
\end{figure}

\newpage

\begin{figure}
\includegraphics[width=0.98\columnwidth]{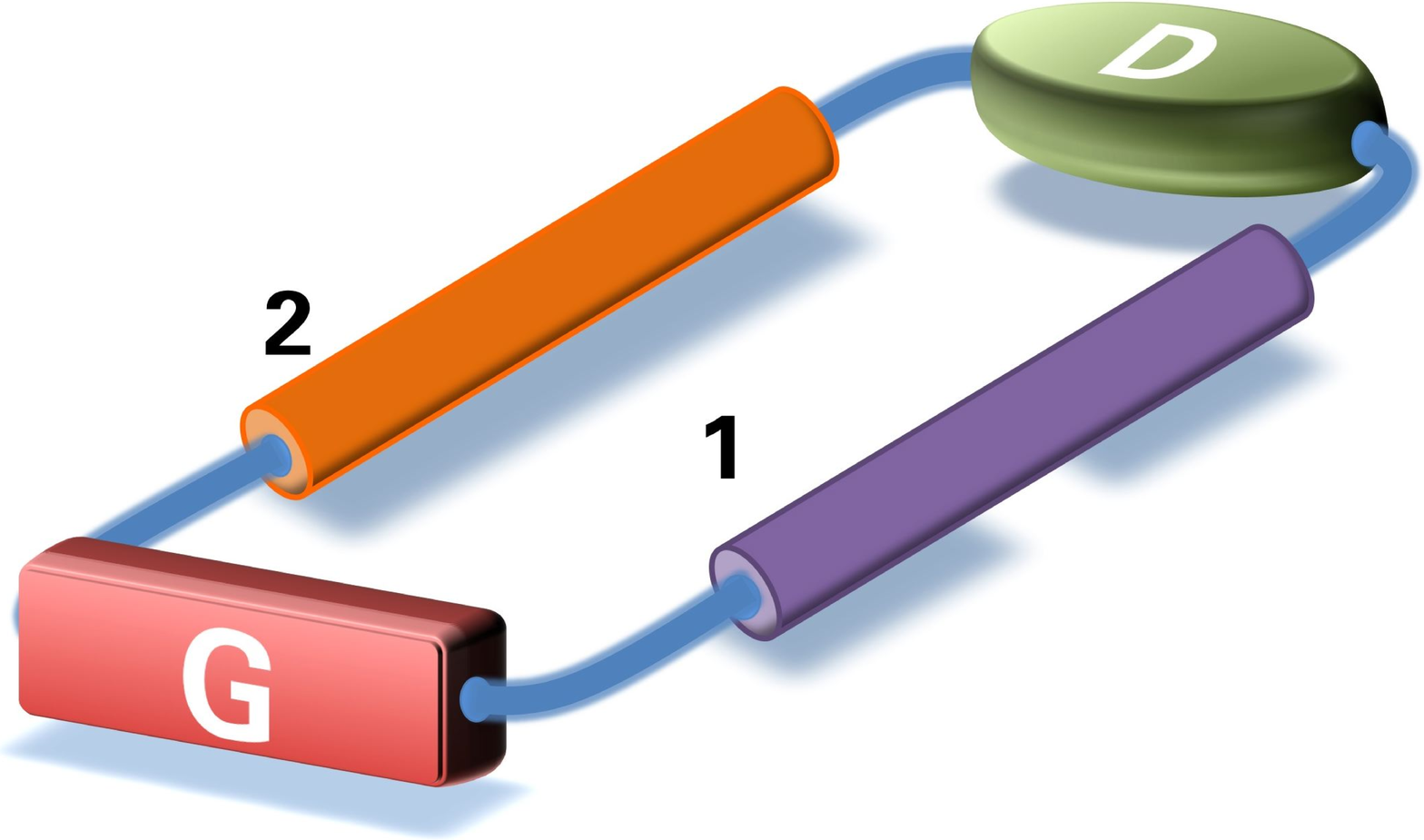}
\caption{}
\end{figure}


\section{References}

\begin{thebibliography}{99}
\bibitem{Gamow1928}  G. A. Gamow, Zeits. f. Physik \textbf{51}, 204 (1928).

\bibitem{Condon1928}  R. W. Gurney and E. U. Condon, Nature \textbf{122, }%
439 (1928) ; Phys. Rev. \textbf{33, }127 (1929) .

\bibitem{Hartmann1962}  T. E. Hartmann, J. Appl. Phys. \textbf{33}, 3427
(1962);


\bibitem{Ciao1997}  R. Y. Chiao and A. M. Steinberg, Tunneling times and
superluminality, Progress in Optics Vol XXXVII (Elsevier, New York, 1997)
Ed. by E. Wolf.

\bibitem{Buttiker2003}  M. B\"{u}ttiker and S. Washburn, Nature \textbf{422}%
, 271 (2003)

\bibitem{Li2001}  C.F. Li and Q. Wang, JOSA B \textbf{18}, 1174 (2001).

\bibitem{Mugnai2000}  D. Mugnai, A. Ranfagni and R. Ruggeri, Phys. Rev.
Lett. \textbf{21}, 4830 (2000).

%

\bibitem{Olkhovsky2004}  V. S. Olkhovskya, E. Recamib and J. Jakielc,
Phys.Rep. \textbf{398, }133 (2004).


\bibitem{Buttiker1988}  M B\"{u}ttiker and R Landauer, J. Phys. C: Solid
State Phys. \textbf{21} 6207 (1988).

\bibitem{Steinberg1993}  A. M. Steinberg, P. G. Kwiat, and R. Y. Chiao,
Phys. Rev. Lett. \textbf{71}, 708 (1993).

\bibitem{Shvartsburg2005}  A. B. Shvartsburg and G. Petite, Eur. Phys. J. D 
\textbf{36}, 111 (2005)

\bibitem{Shvartsburg2007}  A. B. Shvartsburg, V. Kuzmiak and G. Petite,
Phys. Rev. E \textbf{76}, 1127 (2006).

\bibitem{Nanocoatings}  J. B. Pendry and D. R. Smith, Phys. Today \textbf{57}%
, 37 2004 .

\bibitem{Wangsness1986}  R. K. Wangsness, ''Electromagnetic fields'' 2$^{%
\mathrm{nd}}$ Ed. J. Wiley and Sons, NY (1986).

\bibitem{SSW1997}  L. Stenflo, A. B. Shvartsburg and J.Weiland, Contrib.
Plasma Phys. \textbf{37,} 393, (1997).


\bibitem{Ziolkowsky2001}  R.W.Ziolkowsky, Phys. Rev. E \textbf{63}, 046604
(2001).

\bibitem{Nimtz2003}  G. Nimtz, Prog. Quant. Electron. \textbf{27}, 417
(2003).
\end{thebibliography}
\end{document}